# Small anode ion source

V Dudnikov[1], G Dudnikova[2] and A Dudnikov[3]

[1]Muons, Inc, Batavia 6510, IL, USA
[2]Institute of Computational Mathematics and Mathematical Geophysics SB RAS,
 Lavrent'eva Pr. 6, Novosibirsk, 630090, RF
[3]Institute of Laser Physics SB RAS, Lavrent'eva Pr. 15B, Novosibirsk, 630090, RF

E-mail: Dvg43@yahoo.com

**Abstract**. Ion source modification is proposed for efficient production of ion beam and extending of operating Lifetime. Ionization efficiency of the Bernas type ion source has been improved by using a small anode-thin rod, oriented along the magnetic field. The transverse electric field of small anode transport plasma by drift in crossed field to the emission slit. Optimization of the cathode material recycling is used to increase the operating lifetime. Optimization of the wall potential is used for suppression of flakes formation. A three-electrode extraction system was optimized for low energy beam production and efficient space charge neutralization. An ion beam with emission current density up to 60 mA/cm$^2$ has been extracted from discharge in $BF_3$ gas.

## 1. Introduction

A review of ion sources development for ion implantation and isotope separation was presented in book [1]. The Bernas- White ion source (BWIS) most often used recently in high current ion implanters and several version of BWIS manufactured by some industrial companies. Complicated versions of BWIS with two filaments or indirectly heated cathodes used for high current and multicharged ion beam production [2,3]. Further improve of ion beam parameters and increase of the ion source lifetime is necessary for advanced ion implanters.

## 2. Small Anode Source: Discharge configuration

Ion source modification is proposed for efficient production of ion beam and extending of operating Lifetime [4,5]. Ionization efficiency of the Bernas type ion source has been improved by using a small anode-thin rod, oriented along the magnetic field. Design of SAS ion source is shown in Fig. 1. It consist of anode body 2, cathodes 1, 4, cathode insulators 6, emission slit 3, small anode 7, small anode insulators 8, insulator screens 9. A discharge is supported by voltage applied between cathode 1.4 and small anode 7. It was made from 2.4 mm diameter tungsten wire (as filaments) of 50 mm length along discharge chamber supported by ceramic (filaments) insulators 8 in the middle part of the left side- wall of discharge chamber. In the further experiments the SA supporting insulators were shielded by disks 9 for prevention from deposition by filaments vapor. The transverse electric field of small anode transport plasma by drift in crossed field to the emission slit 3. Optimization of the cathode material recycling is used to increase the operating lifetime. Optimization of the wall potential is used for suppression of flakes formation. A three-electrode extraction system as shown in Fig. 2a, was optimized for low energy beam production and efficient space charge neutralization. The calculations and optimization of the extraction geometry were done using the relaxation code PBGUNS developed by Jack Boers of Thunderbird Simulations, Texas. This code, which has been well tested, simultaneously relaxes the shape of the meniscus at the plasma boundary to include the effects of plasma density together with the effects of space charge within the beam as the extracted ions accelerate through the extraction region. Space charge neutralization at the 99% level is assumed beyond the suppresser electrode. Extracting electrodes could be precision moving for optimization of beam formation at different energies. Dependence of analyzed $^{11}B+$ beam current on discharge current at different beam energies are presented in Fig. 2 b. Close solution is proposed in [6].

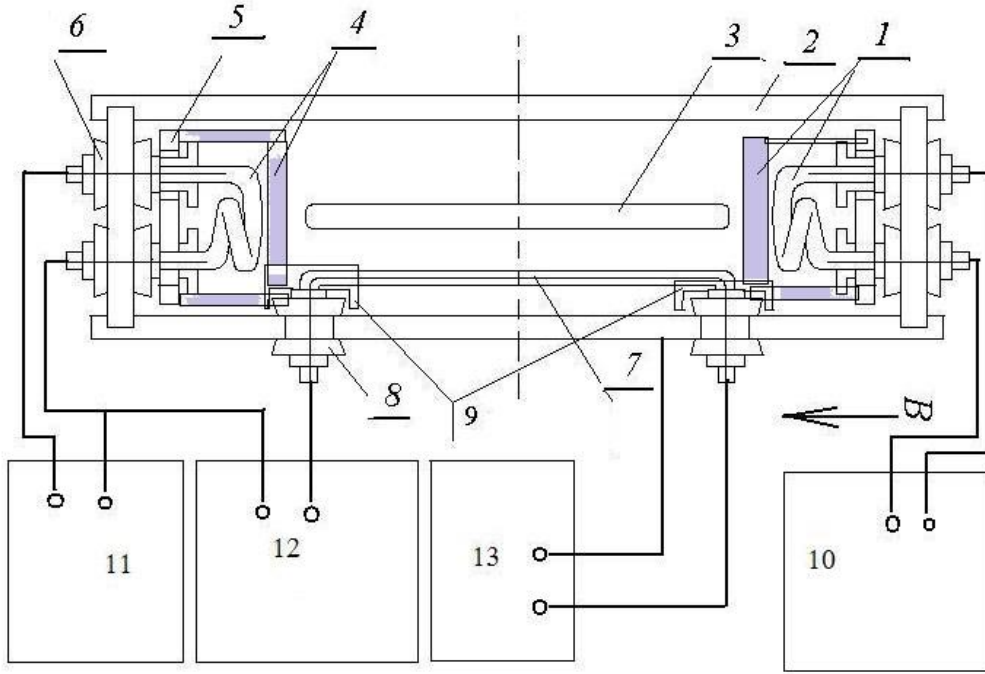

Fig. 1. Schematic of small anode ion source and power supplies.

An ion beam with emission current density up to 60 mA/cm$^2$ has been extracted from discharge in BF$_3$ gas. Ion beams of $^{11}B$ isotope with intensity up to 6 mA for 3 keV, up to 11 mA for 5 keV, 16 mA at 10 keV, 18 mA for 15 keV have been transported through the analyzer magnet as shown in Fig. 2 b) of experimental implanter shown in Fig. 3.

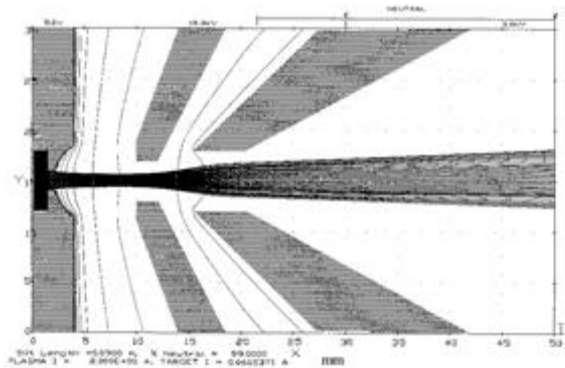 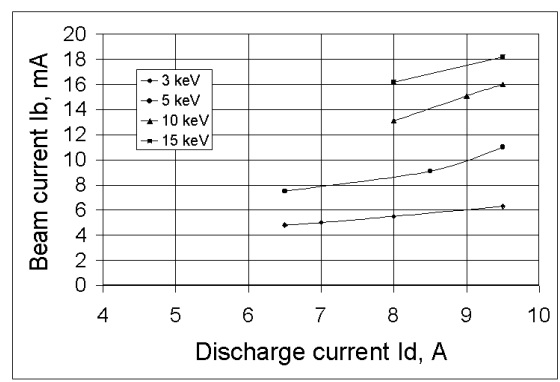

(a)  (b)

Fig. 2. a) Schematic of extraction system of small anode low energy ion source, b) Dependence of analyzed $^{11}B+$ beam current on discharge current at different beam energies.

It consist of ion source12, with extractor 14, analyzing magnet 16, scanning magnet 18, collimating magnet 20, wafer holder 24. After ion Source 12, analyzer magnet 16 analyses the ion beam by removing undesired impurities according to the ion momentum to charge ratio (MV/Q, where v is the velocity of the ion, Q is its charge, and M is its mass). Scanner magnet 18 then Scans the ion beam in a direction perpendicular to the path of the beam. Following Scanning, collimator magnet 20 reorients the ion beam such that the beam is parallel in the entire Scan area. Now similar implanter (Purion) produced by Axcelis .

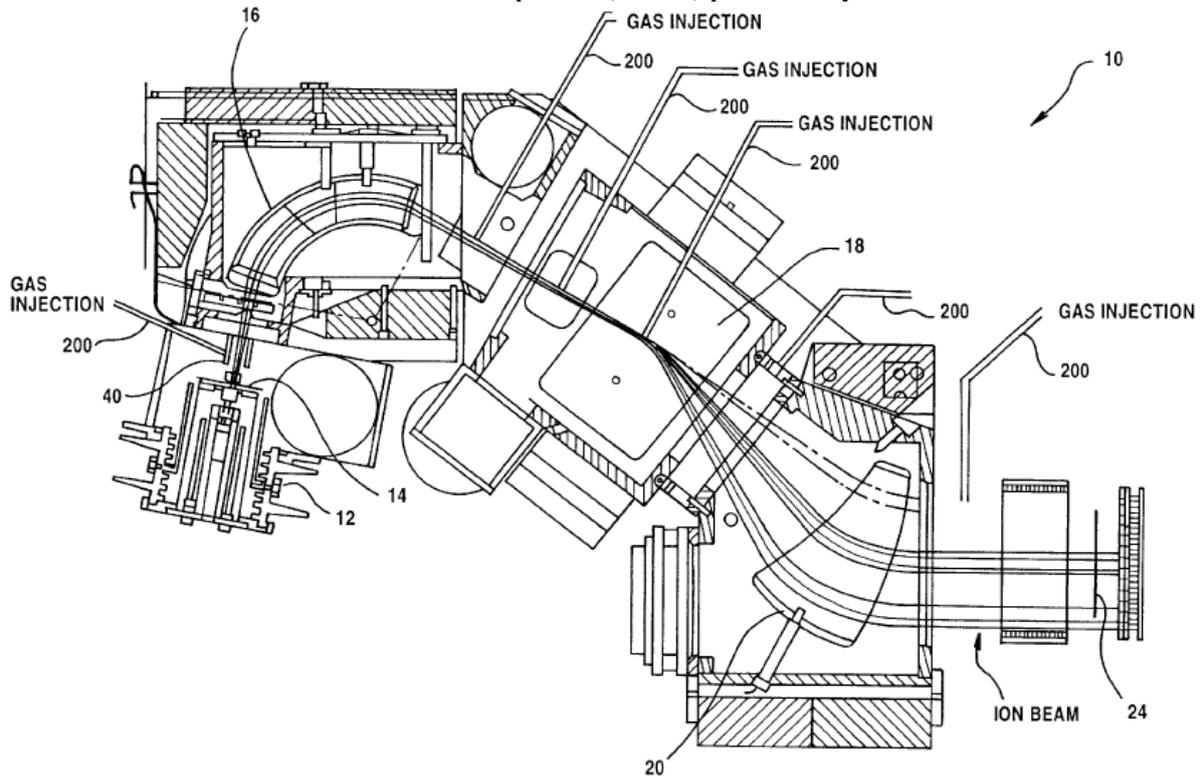

Fig. 3. Schematic of experimental implanter for testing a small anode ion source.

Dependence of $^{11}B+$ current on the discharge current for small anode ion source and for Bernas are presented in Fig. 4 a). Intensity of Small anode source is ~2 times higher then from the Bernas source.

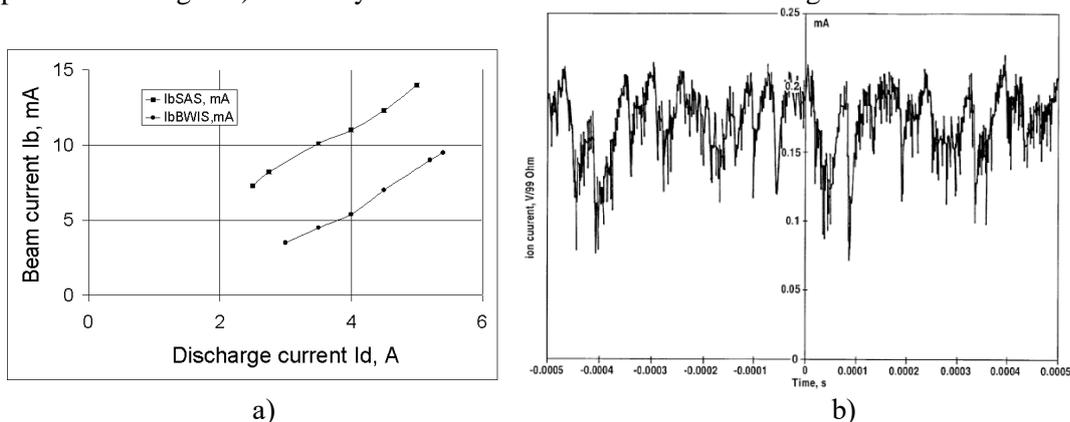

a) b)

Fig. 4. a) Dependences of separated beam $^{11}B+$ ions on arc current for small anode and Bernas ion source. b) Instability of separated beam.

## 3 Space charge compensation of positive ion beam by negative ions

Space charge compensation of positive ion beam by negative ions and stabilization of beam plasma instability by negative ions is described in [7,8,9,10,11]. In almost all previous investigations of SCN of positive ion beams it has been assumed that the compensating particles are electrons. However, in the environment of isotope separation and ion implantation where the complex halide and hydride molecules with high electron affinity are often used as working gases, there is a high probability of negative ion formation. In this situation SCN by negative ions could be significant. Indeed, the SCN by negative ion could be the determining factor for productive operation of large scale ion beam industry, but so far this circumference has not been investigated.

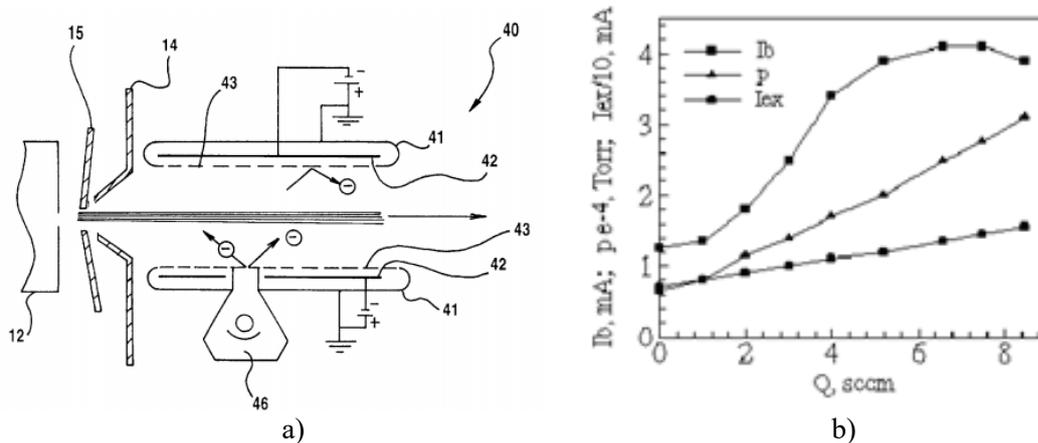

a)                                                                 b)

Fig. 5. a) Embodiment of a gas injection tunnel 40 which increases the density of the electronegative gas near the beam and reflect electrons and negative ions into the beam. b) Dependences of analyzed beam current, extraction current and as pressure on $CF_3$ puffin gas flow.

A possibility to improve SCN by negative ions was tested in beam line shown in Figs. 3 with a magnetic scanner and magnetic suppression of secondary electrons in the beam collector after mass analyzer. For production of high perveance ion beams an ion source with two filaments and small anode made from W wire has been used. Three electrode extractors with precise moving electrodes have been used for beam formation. For low energy beam extraction a high voltage on the suppression electrode (up to -20 kV for 3 keV) was used. Production of high energy neutrals and negative ions in the extractor gap and on the suppressor surface is important for enhance of residual gas ionization and improved space charge neutralization. In a "standard" mode of operation with a strong acceleration-deceleration, low gas density and low noise of discharge in ion source are typically a strong instability of ion beam after analyzer as shown in Fig.4 b). It is typically a relaxation type of oscillation. Beam intensity increases up to critical level, this drives instability, loss of SCN, drop of intensity and then this cycle repeats. It was observed, that with electronegative gases in ion sources, such as $BF_3$, $CF_4$ it is possible to suppress beam instability by increasing gas injection into the source. This instability was not damped by increase of noble gas density such as Ar, or Kr. Improved SCN and damping of instabilities could be related to adding of negative ions into the beam instead of free electrons.

For enhance negative ion formation in ion beams an injection of electronegative heavy molecules with high electron affinity into the ion beam is proposed. Negative ions in ion beam are formed by collision of electrons with molecules and by bombardment of electrodes surfaces by beam and plasma particles. With the use of negative ion for positive ion beam SCN it is possible to create overneutralization, as in the negative ion beam with SCN by positive ions. For low energy ion beam overneutralization discharge plasma and electronegative heavy gas molecules are injected into the beam..

A typical beam line for ion beam production, formation, transportation, separation, scanning, collimation and utilizing consist of the ion source, extraction system, analyzer magnet with mass resolving system, scanner magnet, collimator magnet and end station for the material processing by ion beam. Very good space charge neutralization is necessary in the all parts of ion beam transportation. The strongest space charge forces defocused beam directly after extractor because of multicomponent, high perveance beam is extracted from ion source plasma. The intensity of one component ion beam after analyzer could be considerably lower, but the space charge neutralization of this beam is also important for prevention of the loss of beam intensity and quality. For the prevention of the compensation particles extraction from the beam to the ion source by the electric field of extraction voltage is used a suppression electrode with negative voltage between ion source and grounded extraction electrode reflecting compensating particles into the ion beam.

Negative ions can be introduced into the path of the beam in various location along the path of the beam. Specifically, in ion implanter (Fig. 3), an electronegative gas can be injected, via gas conduits 200, after extraction electrode 14 and before analyzer magnet 16, after analyzer magnet 16 and before Scanner magnet 18, and after Scanner magnet 18 and before collimator magnet 20. The electronegative gas may also be injected in the cavities defined by the analyzer, Scanner, or collimator magnets through which the ion beam travels. The electronegative gas can also be injected after the collimator magnet or just before the beam contacts the wafer.

Fig. 5 a) shows an embodiment of a gas injection tunnel 40 which increases the density of the electronegative gas near the beam. Hence, tunnel 40 also reduces a rate of injection of the gas which would be required to maintain Sufficient gas density for generating a Sufficient density of negative ions. Gas injection tunnel 40 has three parts: an outer wall 41, an electrode 42, and an inner mesh Screen 43. Outer wall 41 is grounded So as to prevent Stray electrical fields from interfering with the beam. Electrode 42 is negatively charged to reflect negative ions and electrons back into the path of the beam. Inner mesh Screen 43 has a Selected degree of transparency, preferably in the order of 90% transparency. A gas tube 44 carries an electronegative gas to a nozzle 45 for injection into tunnel 40. When the electronegative gas is injected into tunnel 40, the gas disperses in the vacuum. Inner mesh Screen 43 reflects Some of the, negative ions and electrons that move away from the beam back into the path of the beam and thereby increases the gas density near the beam relative to other parts of the implanter.

Ion beam neutralization by negative ions is the most important for low energy beams because the cross section of electron production during gas ionization by low energy heavy ions is very low, really zero. An effect of electronegative gas admixture to the 3 keV beam of $B^+$ demonstrated in Fig.5b). Ions of $^{10}B^+$, $^{11}B^+$, $F^+$, $BF^+$, $BF_2^+$,.. was extracted from the 2x90 mm$^2$ slit of ion source with discharge in the $BF_3$ gas. Separated beam of $^{11}B^+$ was registered after 14x80 mass- slit of analyzer magnet by magnetically suppressed collector. Optimized fluxes of electronegative $CF_4$ gas was injected into the tunnel around the beam after extractor and after analyzer. Ion beam current of 3 keV $^{11}B^+$ versus full flux of electronegative gas presented in Fig. 5b).

With increase of gas flux the beam intensity increases from 1.3 mA up to 4.2 mA. With increase of gas density an improvement of focusing by neutralizing the repulsing space charge force and attenuation of beam by charge exchange loss of ions is observed. For real improvement a beam transportation-separation electronegative gas should have a high probability of negative ion formation but low cross section for charge exchange for ions of beam. For $B^+$ ions a good results was reached with $CF_4$, $CClF_3$, $BF_3$ gases. . With $SF_6$ gas in first experiments was observed only beam attenuation but in further experiments with a purified $SF_6$ was observed an increase of intensity. Charge exchange cross section can be large for low energy beam if the fast and slow particles have close ionization potentials (quasi resonant charge exchange). An oscillogram on Fig.6. a) shows the increase of beam intensity after analyzer during increase of the electronegative gas injection. Beam of $^{11}B+$ ions with energy 3 keV was transported and $BF_3$ neutralizing gas with flux up to 4.6 sccm was injected.

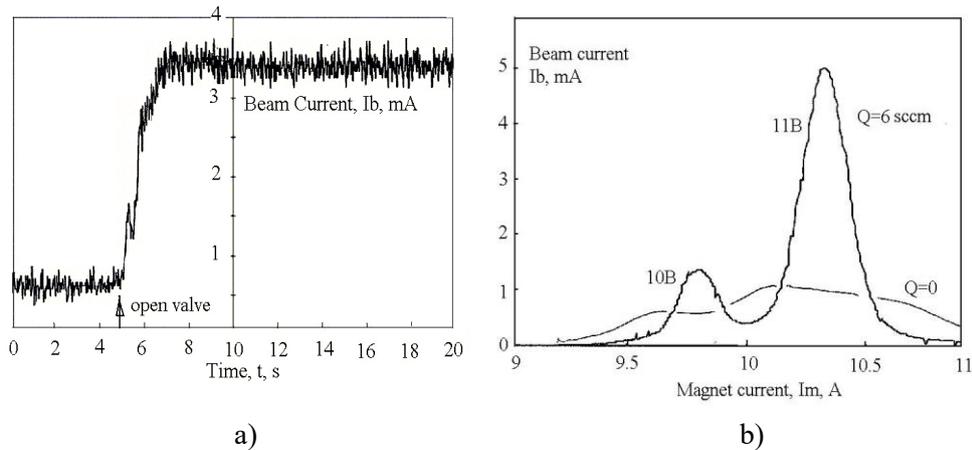

a)                      b)

Fig. 6. a) Oscillogram the increase of beam intensity after analyzer during increase of the electronegative gas injection. Beam of $^{11}B+$ ions with energy 3 keV was transported and $BF_3$ neutralizing gas with flux up to 4.6 sccm was injected. b) Mass spectrum of 3 keV $B^+$ beam after analyser magnet for different flux of neutralizing electronegative gas.

An improvement of low energy (3 keV) $^{11}B^+$ beam by injection of electronegative gas shown on Fig.6 b). Profiles of ion current density distribution were registered after magnetic analyzer. Ion beam current density is increased significantly and beam transverse size decrease with admixture and optimization of the electronegative gas flux. Ion beam instability was dumped by admixture of optimized electronegative gas density as shown on Fig.7.

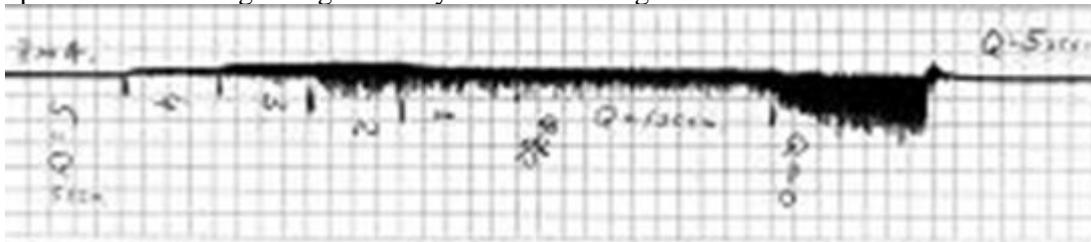

Fig. 7. Ion beam current after analyzer with dumping of the beam instability by injection of electronegative gases.

### 4. Conclusion

Small Anode Source (SAS) with a small, magnetoinsulated anode has been developed and tested. Improved efficiency of ion beam production has been demonstrated. Up to 18 mA of $^{11}B^+$ has been attained behind a mass analyzer. Low level of beam current oscillations has been demonstrated. Increased operation time has been demonstrated. Recycling of cathode material and solid deposition of cathode material without flakes formation has been observed.

A transportation of low energy beam of $B^+$ ion in high current implanters is complicated by very big space charge of molecular ions extracted from discharge in the $BF_3$ gas. A beam intensity and stability have been significantly improved by injection into the beam small admixture of electronegative gases, as $BF_3$, $SF_6$, $CF_4$,…